# Electronic Structure and Chemical Bonding of Amorphous Chromium Carbide Thin Films

Martin Magnuson[1], Matilda Andersson[2], Jun Lu[1], Lars Hultman[1] and Ulf Jansson[2]

[1]*Thin Film Physics Division, Department of Physics, Chemistry and Biology (IFM), Linköping University, SE-58183 Linköping, Sweden*

[2]*Department of Chemistry - Ångström, The Ångström Laboratory, Uppsala University, Box 538, S-75121 Uppsala, Sweden*

Email: Martin.Magnuson@ifm.liu.se

**Abstract**
The microstructure, electronic structure, and chemical bonding of chromium carbide thin films with different carbon contents have been investigated with high-resolution transmission electron microscopy, electron energy loss spectroscopy and soft x-ray absorption-emission spectroscopies. Most of the films can be described as amorphous nanocomposites with non-crystalline $CrC_x$ in an amorphous carbon matrix. At high carbon contents, graphene-like structures are formed in the amorphous carbon matrix. At 47 at% carbon content, randomly oriented nanocrystallites are formed creating a complex microstructure of three components. The soft x-ray absorption-emission study shows additional peak structures exhibiting non-octahedral coordination and bonding.

## 1. Introduction

Metal carbide coatings have many interesting electrical, mechanical, and tribological properties. Examples of multifunctional applications for these materials are combinations of oxidation, wear and corrosion resistance, decorative purposes, and electrical contacts [1,2]. Magnetron sputtering of metal carbide films can give different microstructures ranging from epitaxial single-crystal materials to nanocomposites with crystallites of the carbide in an amorphous matrix [3,4]. It is also possible to deposit completely amorphous metal carbide coatings in both binary and ternary systems such as Cr-C, Fe-C, and W-Fe-C [1]. The explanation for the formation of non-crystalline phases during sputtering in these carbide systems remains unclear, but it is known that amorphous films often are found in carbide systems where the crystalline phases exhibits structural units that have carbon atoms in *both* octahedral and prismatic sites [5].

Magnetron sputtering of chromium carbides including the formation of amorphous Cr-C films has been reported (see, e.g., refs. [6-9]). In a recent study, Andersson *et al.* [10] related the crystallinity of Cr-C coatings to the deposition process used. A survey of the literature showed that reactive deposition tends to give more crystalline films while non-reactive sputtering from elemental targets favors amorphous growth. An observation in Ref. [10] is that elemental target sputtering can give completely amorphous films with two separate phases: i) an amorphous chromium carbide, $CrC_x$ and ii) an amorphous C-rich phase dominated by C-C bonds. The amorphous metal carbide phase contained 20-30 at% of carbon giving a composition between the crystalline $Cr_{23}C_6$ and $Cr_7C_3$ phases while the amorphous carbon phase contained mainly pure carbon with a high amount of *sp²*-hybridized carbon [10]. However, it was also concluded that more studies than x-ray photoelectron spectroscopy (XPS) analysis are required to determine structure and bonding of the two amorphous phases in these films.

In this work, we have studied the detailed chemical bonding and structure of magnetron sputtered amorphous Cr-C films using a combination of high-resolution transmission electron microscopy (HRTEM), electron energy loss spectroscopy (EELS), soft x-ray absorption (SXA) and soft x-ray emission spectroscopy (SXE). An important part of the study is to investigate short- and long-range ordering in the materials and how this is influenced by composition and charge-transfer between the elements. We confirm that the films consist of two amorphous phases with the exception of composition at 47 at% C, where nanocrystallites are formed. The importance of these structures and their interfaces for the film properties is also discussed. These short to medium-range structures and the charge transfer between the elements are known to affect properties such as hardness, elasticity, and electrical resistance.

## 2. Experimental

Four chromium carbide ($Cr_{1-x}C_x$) samples were deposited in Ar atmosphere at 0.4 Pa using non-reactive DC magnetron sputtering in an UHV chamber with base pressure of $1*10^{-7}$ Pa. Two separate 2-inch targets with a purity of 99.999% for C and 99.95% for Cr were used. The films were deposited onto $SiO_2$ substrates at 300 °C and with a substrate bias of -50 V. The distance between targets and substrate was about 14 cm. By tuning the Cr magnetron while keeping the C magnetron constant, compositions between 25 and 85 at.% were achieved at a deposition rate of 12-45 Å/min depending on composition. The compositions of the $Cr_{1-x}C_x$ films were determined using XPS as described in Ref. 10. The deposition of the amorphous carbon (a-C) film is described in Ref [11].

The $Cr_{1-x}C_x$ samples were characterized with an analytical high resolution (HR)-TEM FEI field emission gun instrument operated at 200 kV with 1.95 Å point resolution. The microstructure was analyzed by selected area electron diffraction (SAED) in the TEM equipped with an electron energy loss (EELS) spectrometer with an energy resolution of 0.7 eV. The $Cr_{1-x}C_x$ films were also measured with electron energy loss spectroscopy (EELS) at the C *1s*-edge. The radial distribution functions and bond distances were extracted using the *ATHENA* software package [12] in Fig. 2. The samples were prepared in a conventional way including grinding, polishing and finally ion milling.

Soft x-ray absorption (SXA) and soft x-ray emission (SXE) measurements were performed at the undulator beamline I511-3 at MAX II (MAX-IV Laboratory, Lund, Sweden), comprising a 49-pole undulator and a modified SX-700 plane grating monochromator [13]. The SXA spectra were measured at 5° grazing incidence angle that was found to be necessary for amorphous samples in total fluorescence yield (TFY) mode with 0.15 and 0.1 eV resolution at the Cr *2p* and C *1s* absorption edges, respectively. The SXE spectra were measured with a high-resolution Rowland-mount grazing-incidence grating spectrometer with a two-dimensional detector. The Cr *L* and C *K* SXE spectra were recorded using a spherical grating with 1200 lines/mm of 5 m radius in the first order of diffraction. During the Cr *L* and C *K* SXE measurements, the resolutions of the beamline monochromator were 0.68, and 0.2 eV, respectively.

The SXE spectra were recorded with spectrometer resolutions of 0.7 and 0.2 eV, respectively. The measurements were performed with a base pressure lower than $6.7*10^{-7}$ Pa. In order to minimize self-absorption effects [14], the angle of incidence was 15º from the surface plane during the SXE measurements. The x-ray photons were detected parallel to the polarization vector of the incoming beam in order to minimize elastic scattering. The Cr $L_{2,3}$ and C $K$ SXE spectra were normalized to the main peak heights and were plotted on a photon energy scale (bottom) and an energy scale relative to the $E_F$ (top) in Fig. 3 and Fig. 4.

## 3. Results

A series of Cr-C samples with different compositions were deposited at 300 ºC. The compositions were determined by XPS to $Cr_{0.75}C_{0.25}$, $Cr_{0.53}C_{0.47}$, $Cr_{0.33}C_{0.67}$, and $Cr_{0.15}C_{0.85}$. The compositions are similar to those used in ref. [10].

### 3.1 Transmission electron microscopy

Figure 1 shows HRTEM images at low to high resolutions with corresponding SAED patterns of the $Cr_{1-x}C_x$ samples. As observed, the structural ordering largely depends on the carbon content. The most Cr-rich film in the top panel (x=0.25), is fully amorphous with no indication of (crystalline) ordering. Its SAED pattern shows only weak and broad rings typical for amorphous samples [15]. In contrast, an increase of the carbon content to x=0.47 shows short- and medium-range order within defect-rich nanocrystalline grains with lattice fringe spacing between 1.9 and 2.1 Å in different directions for each crystallite. The SAED pattern from this sample has sharper rings typical for the formation of a crystalline structure with a slightly larger structural distance of 2.21 Å in comparison to the other films. While it is difficult to identify the phase compositions of the randomly distributed grains, these are likely cubic CrC possibly mixed with $Cr_3C_2$. The small white spots in the SAED originate from the substrate.

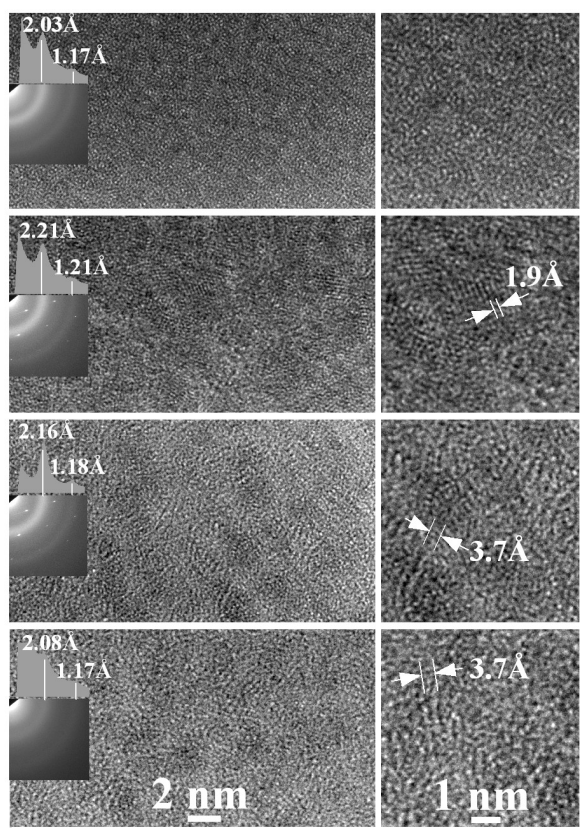

**Fig. 1:** HRTEM images of the $Cr_{1-x}C_x$ thin samples with increasing nominal C content from x=0.25 (top), x=0.47, x=0.67 to x=0.85 (bottom). The corresponding SAED patterns are shown on the left and the HRTEM images in the right panels have double the magnification.

An increase in carbon content to x=0.67 and x=0.85 changes the structure again. No indication of crystalline phases can be seen in these two C-rich samples. However, the HRTEM images in the left panels show dark and light areas. This can be explained by elemental contrast where the darker areas are amorphous domains richer in Cr while the lighter areas can be attributed to amorphous domains more rich in carbon. Higher magnification of the two most C-rich films shows small, embedded carbon clusters with a graphene-like structure with atomic distance between 3.5 and 4.0 Å, clearly observed along the extension of the thin white lines that are guides for the eyes. The amount of graphene layers is not large enough and the small size of these clusters (typically 1 to 10 nm in length and 2 to 4 atomic layers thick) explains why the intensity is too low in the background of the central beam to be observed in SAED.

### 3.2 Radial distribution functions

In order to determine the different bond lengths in the samples, atomic radial distribution functions (RDFs) were extracted by the standard EXAFS procedure [12] from extended electron energy loss fine structure (EXELFS) measurements and the results are shown in Fig. 2. The first peak (1) at the position of ~1.5 Å corresponds to the C-C bond length. For comparison, the C-C bond length is 1.44 Å in amorphous carbon, 1.43 Å in graphite and 1.54 Å in diamond [15-17]. The results qualitatively agree well with the relative distribution of carbon in the carbidic, $CrC_x$ phase and the a-C phase as reported in ref. [10] and summarized in Table 1. We estimate the error bar for the bond length determination within a half standard deviation ±0.15 Å. However, quantitatively, the ratio between the C-C (1) and C-Cr (2) peaks in Fig. 2 is lower in the RDFs than for XPS. As observed, the first peak (1) gradually decreases in intensity with reduced C-content. This signifies that the C-C bonds in free carbon in the films disappear for x=0.25 while XPS shows 11% of the amorphous carbon phase. The second peak (2) at ~2.2 Å is due to the C-Cr bond distance consistent with Cr *1s* EXAFS studies (2.18 Å) [18]. The third (3) and fourth (4) peaks are related the second and third coordination shell of C-C bonding. The experimentally observed C-C (~1.5 Å) and C-Cr (~2.2 Å) bond lengths are shorter than calculated bond lengths for the CrC crystallites; $Cr_3C_2$ (C-C=2.11 Å, C-Cr=2.63 Å), $Cr_7C_3$ (C-C=2.14 Å, C-Cr=2.58 Å) and $Cr_{23}C_6$ (C-C=2.10 Å, C-Cr=2.56 Å) [19]. The RDFs also show structures around ~3.5 Å that could correspond to graphene-like structures but the uncertainties with overlapping high order coordination shells are too large to draw any definite conclusions.

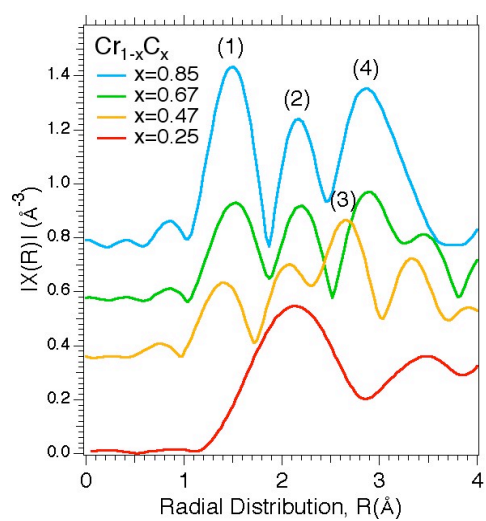

**Fig. 2:** (Color online) Radial distribution functions (RDFs) of $Cr_{1-x}C_x$ extracted from EELS at the C *1s* absorption edge.

**Table I:** Composition of the $Cr_{1-x}C_x$ films for x=0.25, 0.47, 0.67 and 0.85. The C-C and C-Cr chemical bond contributions were determined by integrating the areas under the corresponding peak structures in C *1s* XPS spectra [10]. The Cr $L_3$ SXE peak positions and $L_3/L_2$ ratios in Fig. 3 are also listed.

| Composition (XPS) | $Cr_{0.75}C_{0.25}$ | $Cr_{0.53}C_{0.47}$ | $Cr_{0.33}C_{0.67}$ | $Cr_{0.15}C_{0.85}$ |
|---|---|---|---|---|
| % a-C phase | 11 | 61 | 85 | 98 |
| % $CrC_x$ phase | 89 | 39 | 15 | 2 |
| Cr $L_3$ (eV) (SXE) | 576.38 | 575.98 | 576.48 | 575.79 |
| Cr $L_3/L_2$ (SXE) | 3.46 | 3.02 | 2.88 | 2.40 |
| $\pi^*/[\pi^*+\sigma^*]$ (C *1s* SXA) | 0.074 | 0.093 | 0.079 | 0.095 |

### 3.3 Cr *2p* x-ray absorption spectra

Figure 3 (top) shows Cr *2p* TFY-SXA spectra following the Cr $2p_{3/2,1/2} \rightarrow 3d$ dipole transitions of the $Cr_{1-x}C_x$ films with different carbon content in comparison to Cr metal. Three peaks are observed at 2 eV, 4.2 eV and 5.5 eV on the energy scale at the top of Fig. 3 denoted (1), (2), (3) at the $2p_{3/2}$ threshold and two peaks at 13.2 eV and 14.5 eV at the $2p_{1/2}$ threshold denoted (4), (5). Peaks (2), (3) and (4), (5) are expected and correspond to the usual $t_{2g}$-$e_g$ type of ligand-field splitting in transition-metal carbides while peak (1) is a new feature. A comparison of the spectral shapes at different carbon contents shows two interesting effects: (i) the intensity of the spectra increase with carbon content as in the case of Ti *2p* SXA in nc-TiC/a-C nanocomposites [11]. The Cr *2p* SXA intensity is proportional to the unoccupied *3d* states and we can therefore conclude that the Cr *3d* electron density is reduced around the absorbing Cr atoms for the higher carbon concentrations. The broad low-energy tail most visible for high carbon contents and the high step in intensity from below to above the thresholds is characteristic of the large sensitivity of the fluorescence yield method at low incidence angle. (ii) The SXA spectrum of Cr metal (x=0) has narrower and less intense $2p_{3/2}$ and $2p_{1/2}$ absorption peaks whereas the SXA spectra of the carbon-containing films exhibit additional peak structures over a wider energy range at the $2p_{3/2}$ peak. At the $2p_{1/2}$ threshold, the spectra are further broadened with a different intensity ratio than at the $2p_{3/2}$ threshold due to the exchange-correlation mixing between the core-states [20] and the Coster-Kronig process [21]. In particular, we observe that the *2p* SXA peak structures are different from the other spectra for x=0.47 and x=0.67 with the unexpected and growing peak (1) in comparison to the single $2p_{3/2}$ peak (3) of Cr metal. This is an indication of a change in average coordination for the absorbing Cr atoms with composition as will be discussed below.

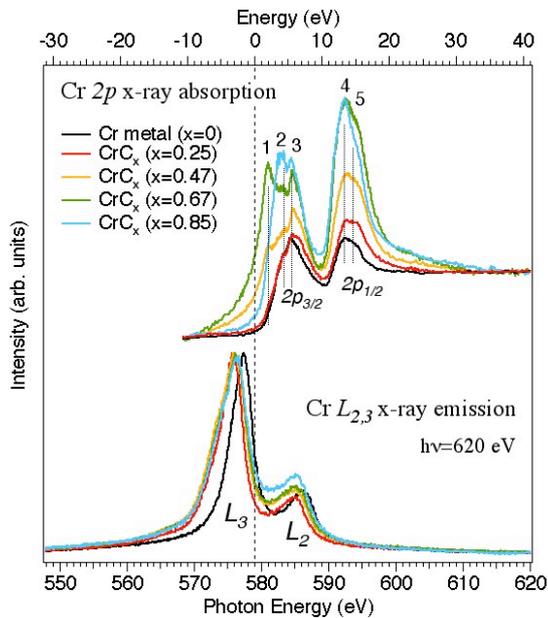

**Fig. 3:** (Color online) Top: Cr *2p* TFY-SXA spectra of $Cr_{1-x}C_x$ with different carbon contents x (in parenthesis) in comparison to bulk Cr metal (x=0). Bottom: Cr $L_{2,3}$ SXE spectra of $Cr_{1-x}C_x$ and Cr metal nonresonantly excited at 620 eV.

### 3.4 Cr $L_{2,3}$ x-ray emission spectra

Fig. 3 (bottom) shows Cr $L_{2,3}$ SXE spectra of $Cr_{1-x}C_x$ in comparison to Cr metal, nonresonantly excited at 620 eV. The SXE spectra probes the occupied Cr *3d* states following the Cr *3d* -> $2p_{3/2,1/2}$ dipole transitions. Starting with the Cr $L_{2,3}$ SXE spectra of Cr metal (x=0, black curve), the main $L_3$ emission line due to metal Cr *3d* - Cr *3d* interactions is observed at -1.5 eV on the energy scale at the top of Fig. 3. The broader and less intense $L_2$ emission line is observed at +7 eV for Cr metal and at +6 eV for the carbide films. For the $Cr_{1-x}C_x$ films there are three interesting observations; (i) the $L_3$ peak is shifted 0.5-1.0 eV to lower energies and, (ii) the $L_3$ peak is significantly broader compared to Cr metal and, (iii) the $L_3/L_2$ branching ratio decreases with increasing carbon content (Table I). For the Cr $L_{2,3}$ SXE spectra of the $Cr_{1-x}C_x$ films, the main $L_3$ peak is due to Cr *3d* - C *2p* hybridization similar to the case of nc-TiC/a-C nanocomposites [11]. The low-energy shifts away from the Fermi level and the increased broadening of the $L_3$ and $L_2$ peaks for increasing carbon content can be attributed to a stronger Cr *3d* - C *2p* interaction compared to the Cr *3d* - Cr *3d* interactions in Cr metal [21]. The trend in the $L_3/L_2$ branching ratio in transition metal compounds is a signature of the degree of ionicity in the systems. For conducting systems, the $L_3/L_2$ ratio is usually significantly higher than the statistical ratio 2:1 due to the additional Coster-Kronig decay process [21]. The observed $L_3/L_2$ ratio systematically decreases as the carbon content increases (Table I), and is a sign of decreased metallicity [22]. Consequently, the Cr atoms in the most C-rich $Cr_{1-x}C_x$ films appear to be more ionic than those with lower carbon content. For Cr metal, the $L_3/L_2$ ratio higher than 3:1 is a metallic signature.

### 3.5 C *1s* x-ray absorption spectra

Figure 4 (top) shows C *1s* TFY-SXA spectra of the $Cr_{1-x}C_x$ samples and amorphous carbon (a-C) [11], where the intensity is proportional to the unoccupied C *2p* states. The intensity of the C *1s* SXA spectra generally follow an opposite trend in comparison to the Cr *2p* SXA spectra with an intensity decrease due to increased electron density around the carbon atoms for increasing carbon content. The first absorption peak at 1.4 eV denoted (1) on the energy scale at the top of Fig. 4 is due to empty π* orbitals and consists of two contributions: (i) $sp^2$ (C=C) and $sp^1$ hybridized C states in the amorphous carbon phase and, (ii) C *2p* - Cr *3d* hybridized states in the amorphous chromium carbide phase. Peak (2) at 5.0 eV is also due to C *2p* - Cr *3d* hybridization with addition of the superimposed carbon phase [11] and possible C-O bond contributions. The energy region above 6 eV is known to originate from $sp^3$ hybridized (C-C) σ* resonances where peak (3) at 7.5 eV forms a broad shape resonance with multielectron excitations towards higher energies [11]. All the π* peaks (1) of the carbides have a 1 eV shift towards higher energy in comparison to a-C. Due to the many overlapping carbon contributions and resonances, the fine structure of the spectra is complicated and it is difficult to distinguish the contribution from each feature to the total spectra. While the intensity of the σ* peak (3) and π* peak (2) follow the expected trend, the π* peak (1) show a slightly different trend. By calculating the integrated $\pi^*/[\pi^* + \sigma^*]$ intensity ratio given in Table I, an estimation of the relative amount of $\pi^*$ content in the samples can be made following the procedure in Ref. [23].

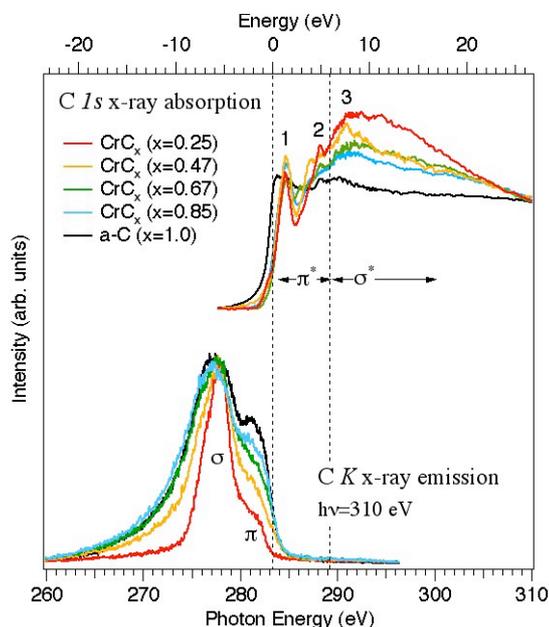

**Fig. 4:** (Color online) Top: C *1s* TFY-SXA spectra of $Cr_{1-x}C_x$ with different carbon contents (in parenthesis) and amorphous carbon (a-C). Bottom: nonresonant C *K* SXE spectra of $Cr_{1-x}C_x$ and a-C excited at 310 eV.

### 3.6 C K x-ray emission spectra

Figure 4 (bottom) shows C K SXE spectra of the $Cr_{1-x}C_x$ films excited at 310 eV photon energy, probing the occupied C *2p* states of the valence bands following the C *2p->1s* dipole transitions with different carbon content in $Cr_{1-x}C_x$ in comparison to a-C. As in the case of the C *1s* SXA spectra, the C K SXE spectra of the $Cr_{1-x}C_x$ films are composed of a superposition of at least two contributions; i) one from the amorphous carbon phase and another from ii) the amorphous chromium carbide phase. The main peak corresponds to the occupied C *2p* σ band at -5.5 eV with a high-energy band at -2 eV with π character. For low carbon content (x=0.25), the dominating σ peak is characteristic of strong covalent Cr *3d* - C *2p* bonding in the chromium-carbide phase. Note that contrary to the case of octahedrally coordinated carbon atoms in TiC [11,24], the low-energy shoulder is not clearly resolved from the main σ peak but is imbedded in the C K SXE spectra at -6.5 eV for the chromium carbide with low carbon content (x=0.25). As the carbon content increases, the imbedded low-energy shoulder appears to gain weight as the σ peak broadens and shifts about 1 eV to lower energy, developing a broad low-energy tail while the intensity of the π-band also increases. A comparison of the relative amount of $sp^2$ hybridization can be made by calculating the measured π/σ peak intensity. As the carbon content increases, the occupied π/σ peak intensity ratio systematically increases from 0.3 to 0.6. For x=0.85, the π/σ intensity ratio of 0.7 is almost as high as for amorphous carbon (a-C) and thus has the most $sp^2$ hybridized content due to the graphene-like structures observed in the HRTEM that occur in the samples for high carbon content. However, the intensity of the π band at -2 eV is higher for x=0.85 than in graphite [23], indicating additional contribution of $sp^3$ hybridized C at the top of the valence band as in the case of diamond that is completely $sp^3$ hybridized [25].

### 4. Discussion

The HRTEM images in Fig. 1 show that the $Cr_{1-x}C_x$ films are mainly amorphous. The most chromium-rich film $Cr_{0.75}C_{0.25}$, as well as the most C-rich films $Cr_{0.33}C_{0.67}$ and $Cr_{0.15}C_{0.85}$ show no indication of crystallinity. This confirms the observations in Ref. [10] based on x-ray diffraction (XRD) and XPS that the films consist of two amorphous phases; i) a chromium-rich carbide $CrC_x$ phase and, ii) a C-rich phase. The presence of two amorphous phases is also supported by the C-C and C-Cr RDF peaks in Fig. 2 with characteristic atomic distances. The SAED and RDF from EELS provide complementary information due to the very high sensitivity of the loss energies in EELS for light elements such as C compared to SAED that is more sensitive to heavier elements such as Cr. The observation of areas with graphene-like structures in the most C-rich films suggests a considerable $sp^2$ hybridization of carbon. This is also supported by Raman data of similar films reported in ref. [10] that indicate that the *sp²*-contribution in the a-C phase is very high (≥80%). The graphene-like structures in Fig. 1 consist of several layers (2-4) of carbon with a layer distance of 3-4 Å. It is likely that these graphene-like structures influence some of the properties but are too rare to be observed in spectroscopy. The presence of two phases is also supported in Fig. 2 that shows C-C bonds and C-Cr bonds. The variations of the intensities are qualitatively similar with the variations observed in XPS.

The $Cr_{0.53}C_{0.47}$ film exhibits a different microstructure compared to the other films. In this case, we observe randomly oriented nanocrystallites embedded in a matrix of both amorphous carbon as well as non-crystallized amorphous $CrC_x$. Thus, this sample consists of three phases; two amorphous phases and one nanocrystalline chromium carbide phase. As observed in Fig. 1, the nanocrystallites have a typical size of less than 3 nm with different shapes and orientation. XRD data published in Ref. [10] showed no sign of crystallinity for this composition that can be explained by the small grain size making the films x-ray amorphous. The nanocrystallites in the $Cr_{0.53}C_{0.47}$ sample have crystalline plane distances in the range 1.9 Å to 2.1 Å. Several chromium carbides such as $Cr_3C_2$ and metastable CrC have plane distances in this range and it is therefore difficult to identify the exact crystal structure of the embedded nanocrystallites. The crystallization kinetics of amorphous chromium carbides as a function of composition has not yet been studied and it is therefore not clear why only the $Cr_{0.53}C_{0.47}$ film has a partially crystallized carbide phase. However, it should be noted that this composition is close to the metastable cubic CrC phase observed by Gassner *et al.* [26] and is close to the composition of $Cr_3C_2$. It is reasonable that total film compositions close to those of the crystallization products favor nucleation. However, further studies of the crystallization kinetics are required to confirm this.

Further details of the chemical bonding and the structure in the chromium carbide films are obtained from the SXA and SXE spectra in Figs. 3-4. The intensity of the Cr *2p* SXA peaks in Fig. 3 is clearly increased with increasing carbon content. This reflects a change in the density of unoccupied states and is attributed to a reduced Cr *3d* electron density at higher carbon contents as will be discussed in more detail below. Notably, a characteristic observation in the Cr *2p* SXA spectra is that the intensity and position of the (1), (2), (3) sub-peaks at the *2p₃/₂* absorption threshold is strongly dependent on the carbon content. The total intensity trend for increasing carbon content in the Cr SXA spectra agrees with the Ti *2p* SXA spectra in an earlier study [11]. In contrast to the $Cr_{1-x}C_x$ films, the Ti *2p₃/₂* and *2p₁/₂* SXA peaks in nanocomposite nc-TiC/a-C films were strongly split into two sub-peaks due the $t_{2g}$-$e_g$ symmetry in the cubic TiC phase. The energy positions of the $t_{2g}$-$e_g$ sub-peaks in the Ti SXA spectra were independent on the total carbon content since all films contained cubic TiC with a well-defined nano-crystalline structure. This suggests that the large variations in intensities and energy positions of the (1), (2), (3) sub-peaks in Fig. 3 reflect changes in the bonding and coordination of the Cr atoms in the amorphous $CrC_x$ phase. Peaks (2), (3) and (4), (5) are similar to the $t_{2g}$-$e_g$ splitting observed in the cubic TiC crystallites [11] but considerably weaker. Although we have not identified the specific origin of peak (1), it likely reflects a different coordination and bonding where the specific structure is unknown. The exact coordination and bonding of Cr in the amorphous $CrC_x$ domains is also unknown, but the results in ref. [10] show that the total carbon content is about 20-30 at%. This is in the same range as the crystalline $Cr_{23}C_6$ and $Cr_7C_3$ phases. In these crystalline phases, the C atoms are placed in a mixture of octahedral and trigonal prismatic sites. It is likely that a mixture of such coordination polyhedra exists in the amorphous $CrC_x$ domains and the distribution is dependent on the total carbon content. Additional experimental and theoretical studies are needed to fully correlate the observed features (1), (2), (3) in the Cr *$2p_{3/2}$* peak with structure. The intensity and position of the peaks are changed as a function of composition. For x=0.47 and 0.67, the spectra exhibit a significant intensity increase of peak (1) that may be due to a change of coordination, while for x=0.85, an additional increase of valency of the Cr atoms likely accounts for a shift of peak (1) towards peak (2). Similar spectral changes and shifts have been observed in hard x-ray SXA at the Cr *1s*-edge when going from octahedral to tetrahedral systems [27,28].

For the Cr *$L_{2,3}$* SXE spectra in Fig. 3, we observe a broadening of the spectral shape with increasing carbon content in a similar way as for the Ti *$L_{2,3}$* SXE peaks in the nanocomposite nc-TiC/a-C films [11]. The Cr *$L_3$* peak exhibits a 0.5-1.0 eV low-energy shift for the $Cr_{1-x}C_x$ films in comparison to pure Cr metal. The chemical shift of the Cr *$L_3$* peak towards lower energy systematically increase for increasing carbon content. Simultaneously, the *$L_3/L_2$* ratio systematically decreases as the carbon content increases (Table I) signifying that the Cr atoms in the most C-rich $Cr_{1-x}C_x$ films are less metallic than those with lower carbon content. This is in agreement with the observations of the trend of the *$L_3/L_2$* ratio of Ti in nanocomposite nc-TiC/a-C films [11] and the increasing resistivity of the $Cr_{1-x}C_x$ samples with increasing carbon content [10].

The C *K* SXE spectra in Fig. 4 originate from several different types of carbon. The most important types are i) carbon from the amorphous carbide, $CrC_x$ and from ii) the a-C phase. However, interface bonding between amorphous carbide and a-C domains will likely contribute. For the $Cr_{0.53}C_{0.47}$ sample, an additional contribution comes from carbon in the nanocrystallites. Comparing the spectral shapes of the C *K* SXE spectra with calculated density of states, the weight of the low-energy shoulder depend on the specific structure [19,29,30]. Experimentally, the overlap between the low-energy shoulder and the main σ peak show the complexity of the amorphous structure. The low-energy shoulder successively gain weight relative to the main σ peak as also previously observed in nanocomposite nc-TiC/a-C films [11]. In contrast to C *K* SXE spectra of graphite and diamond [31] where dispersive peaks are due to the symmetry resolved band structure, the spectral shapes of the $Cr_{1-x}C_x$ samples do not exhibit any significant excitation energy dependence.

If we compare the influence of carbon content on the C *1s* SXA spectra, we observe a difference compared to nanocomposite nc-TiC/a-C films in ref. [11] where the intensity of the unoccupied states was observed to systematically decrease for increasing carbon content. In table I, no clear trend can be seen in the π*/[π*+σ*] contributions as a function of C-content. Furthermore, the C *1s* SXA spectra are a sum of contributions from the a-C and a-$CrC_x$ phases, and since the relative amount of these two phases varies with C-content, it is impossible to draw any quantitative conclusions regarding *$sp^2$* content in the amorphous a-C phase.

Another important detail is the effect of charge-transfer between and within the different structures. The relative intensities of the Cr *2p* SXA spectra show that the Cr atoms in the most C-rich samples are most affected by charge transfer from the Cr *3d* to the C *2p* orbitals. The decrease of the occupied Cr *3d* states and the SXE *$L_3/L_2$* branching ratio show that the Cr atoms become more ionic for increased carbon content, an effect of charge transfer. For the C *K* emission spectra, the situation is opposite to that of the Cr spectra and the intensity and the occupation of the C *2p* states systematically increase due to charge transfer for increasing carbon content. The trend in the C *1s* SXA spectra is less obvious, but the least empty C *2p* states are generally observed for the most C-rich samples. However, due to the smaller difference in electronegativity between Cr - C in comparison to Ti - C, we expect the charge transfer effect to be smaller in the former than in the latter system. The charge transfer from Cr to C may occur within the amorphous $CrC_x$ domains, but also at the interfaces between the domains as an overlapping effect in the C *1s* SXA spectra. The size of the charge-transfer depends on the total surface area at the interfaces. It is difficult to estimate the interface area in the different samples as a function of C-content, due to, for example, the formation of nanocrystallites at x=0.47 and the nature of the a-C phase (formation of graphene-like structures). We can therefore not, in contrast to the nc-TiC/a-C case [11] quantify the charge-transfer effect. This calls for further experimental and theoretical studies on the electronic structure of amorphous transition metal carbide materials.

## 5. Conclusions

We have shown that the structure, bonding, and properties of chromium carbide ($Cr_{1-x}C_x$ with x=0.25-0.85) films deposited by magnetron sputtering are very sensitive to the carbon content. While at low carbon content x=0.25, the samples are fully amorphous with no structural order, the amount of short-range ordering and charge transfer increase for higher carbon content. For x=0.47, nanocrystalline grains are observed while for x=0.67, and x=0.85, domains of different carbon content with short- and medium-range structural order are identified. For these compositions, we demonstrate that the films consist of two-phase amorphous domain structures with an amorphous carbon-rich matrix phase and an amorphous chromium-carbide phase. The presence of two amorphous phases likely affects the corrosion properties of this material. For the higher carbon contents, x=0.67 and x=0.85, the coordination of the Cr atoms appears to transform as the valency increases and the films also contain graphene-like layers between amorphous domains and clusters. The Cr *2p* SXA spectra reveal features that are different from the simple octahedral coordination, suggesting a more complex coordination and bonding. Finally, charge transfer effects are expected in the system and an increasing ionicity of the Cr atoms is observed in Cr SXE for increasing carbon content. However, due to the complex nature of the C *1s* SXA spectra with contributions from different types of carbon, this effect is difficult to quantify for each component in the samples.

## 6. Acknowledgements

We would like to thank the staff at the MAX-IV-Laboratory for experimental support. This work was supported by the Swedish Research Council, the Göran Gustafsson Foundation, and the Swedish Foundation for Strategic Research (SSF) through the research programs ProViking and $MS^2E$.